\documentclass[submission,copyright,creativecommons]{eptcs}
\providecommand{\event}{Wivace2013 - Topic: Complex Systems} 
\usepackage{breakurl}             

\title{Meta-Structures: The Search of Coherence in Collective Behaviours (without Physics)}
\author{Gianfranco Minati\textsuperscript{1} \and Ignazio Licata\textsuperscript{2} \and Eliano Pessa\textsuperscript{3}
\institute{
\textsuperscript{1} Italian Systems Society\\ Via Pellegrino Rossi 42, Milan, I-20161, Italy\\
\textsuperscript{2} ISEM, Institute for Scientific Methodology\\via Ugo La Malfa 153, 
Palermo, I-90146, Italy\\ \textsuperscript{3} Department of Brain and Behavioral Sciences Universit\`a di Pavia\\Piazza Botta 11, I-27100 Pavia, Italy}
}

\def\titlerunning{Meta-Structures: The Search of Coherence in Collective Behaviours (without Physics)}
\def\authorrunning{G. Minati, I. Licata, E. Pessa}

\begin{document}
\maketitle

\begin{abstract}
This contribution shortly outlines and reviews a theoretical and computational approach for a \textit{theory of change} concerning systems where it is not possible to apply the laws of motion \textit{ab initio}. The concept of meta-structure relates to the emergence of forms of spatiotemporal coherences in collective behaviours intended as coherent sequences of multiple structures. The essential difference compared with traditional methods is the role of the cognitive design by the observer when identifying multiple mesoscopic variables. The goal is both to study the ``change without physics'' of the dynamics of change and to design non-catastrophic interventions having the purpose to induce, change, keep or restore collective behaviours by \textit{influencing} -at the mesoscopic level- and not \textit{prescribing} explicit rules and changes.\\

\end{abstract}

\section{Introduction}

In most biological, cognitive and socio-economic phenomena we deal with processes of change that can not be described by the laws of motion and energy conservation of classical physics. Significant parts of the mathematical apparatus traditionally used to study the collective dynamics appear useless. Therefore, we need to introduce a distinction between the evolutionary dynamics,``given'' by typical equations of ``ideal'' models, and the dynamics of change requiring ``not ideal'' simulation approaches \cite{Pessa2009, Licata2010a}. Examples of typical dynamics of change are given by the behaviour of flocks, swarms, ecosystems, traffic, markets, social groups and communication processes (see, for a review, Vicsek \& Zafeiris 2012 \cite{Vicsek2012}, Viswanathan  \textit{et al.}, 2011 \cite{Viswanathan2011}; Kerner, 2004 \cite{Kerner2004}).

In these processes the aspects of ``irreducibility'', lacking in ``ideal model'', derive from the fact that the boundaries between a system and its environment vary quickly and in unpredictable ways, making ``fuzzy'' the distinction between these two entities. Furthermore there appear constraints affecting the relationship between local and global components, or agents, so redefining the patterns into play.

The study of these processes requires a plurality of integrated models, each targeted to a specific aspect of the change under consideration \cite{Minati2008, Minati2009}. In this regard there is a wide variety of simulation approaches, such as neural networks or cellular automata, where the prescriptions on the relations between individual components and boundary conditions give rise to processes of emergence comparable to those typical of the observed systems.

These approaches have in common with the ideal models the same ``philosophy of prediction'' traditionally at the basis of theoretical physics. The idea is to be able ``to zip'' the essential characteristics of change in a set of ideal equations, typically a Lagrangian formulation based on general symmetry or conservation principles. After all, such a formulation allowed the building of models of the so-called ‘spontaneous symmetry breaking’ phenomena, used to introduce the first, basic theories of emergence \cite{Anderson}.

Unfortunately, despite years of intensive research, the mapping between ideal and not ideal models is still an open challenge, chiefly when dealing with the complexity characterizing multiple systems or Collective Beings. The former are constituted by individual elements which can, simultaneously or sequentially, belong to different systems. Among the many possible examples of multiple systems we can quote networks of interacting computers engaged in multiple shared tasks as on the Internet, or electronic devices where the state of a component can be simultaneously a `state' for another system, e.g. devoted to control. The Collective Beings (CB), introduced in \cite{Minati2006}, are special cases of multiple systems in which their components are provided with a cognitive system complex enough to enable them to \textit{decide} their roles such as belonging to temporary communities of mobile phone networks, traffic, communities of passengers on transport systems and queues. We use, in general, the term meta-structures to denote the study of sequences of spatiotemporal patterns in processes of change concerning multiple systems \cite{Minati2012a, Minati2012b, Minati2012}.

The complexity of many meta-structures often entails that the processes of change are essentially ``historical'' and irreversible, being strongly influenced by local constraints. Within such a context the prediction attempts based on ``prescriptive models'', relying on abstract first principles, can be widely disappointed. A possible way for overcoming these difficulties can be the one of investigating the change starting from the past, that is, from the observation of its phenomenological history\footnote{This feature of  meta-structures can be related to the research \textit{in physics} on the \textit{emergence of meta-structures}, i.e., structures of structures formed in the case of topological defects like defects or irregularities in the structures, vortices and transition regions (walls) in systems and in long-range correlations \cite{Pessa2012}.}. This does not mean that traditional mathematical models cannot be used. The problem is that within these systems the relationship between the experimental observations and the needs, motivations, cognitive schemata of the observer itself is so tight as to make impossible the introduction of a \textit{single model} and we need to resort to a dynamic usage of a plurality of models. Such a circumstance is common in many domains of scientific research, such as, e.g., the study of phase transitions in condensed matter physics (examples can be found in Gershenson and Fern\'andez, 2012 \cite{Gershenson2012}; Prokopenko \textit{et al}., 2011 \cite{Prokopenko2011}).

We have defined such approach as ``without physics'' because we deal with processes for which it is impossible to start from a global Lagrangian. The study of meta-structures, instead, search for significant ``a posteriori'' correlations within the history of the change itself. The aim is to detect emergence processes, so as to know what influences warrant the appearance of meta-structures and their permanence. In addition to the traditional prediction goals, one tries to manage and design interventions on the process. Such approach is different from the purely statistical ones not so much from a technical point of view but mainly for the central role of the observer's cognitive design activity in choosing the observables' set \cite{Licata2010b}.

Within these contexts it is therefore necessary to choose different set of observables in relation to the systemic perspective used, particularly the emergent acquired properties detected, and the related processes of emergence to be modelled. This approach can be called `mesoscopic' because it focuses on observables able to \textit{grasp} the system / environment redefinition around the emergence of the local constraints which change the relations between the elements. Namely it deals with phenomena occurring on a scale which neither the one of single components (microscopic) nor the one of a infinite number of components (macroscopic).

The mathematical essential ingredient is the construction of a set of non equivalent, mesoscopic general vectors. Each mesoscopic vector is calibrated on the aspects of the multiple system under consideration with a suitable choice of observables. The observed evolutionary patterns of the mesoscopic vectors can, then, be analysed though suitable data mining and statistical tools in order to detect specific emergence processes and, eventually, help to choose the best non-ideal models describing them.

\section{Modelling and representing mesoscopic dynamics}

The simplest description of multiple systems is based on mesoscopic vectors whose components -- the \textit{mesoscopic variables} -- are obtained through a suitable coarse graining procedure. The latter, for each variable characterizing the single system elements, starts from all differences between the values of this variable, observed at a given time instant \textit{t}, in all possible element pairs and, resorting to suitable threshold values, sets equal to zero all differences whose value is lesser than the threshold. This allows to define a new mesoscopic variable, given by the number of elements having the same value of the observed variable. These mesoscopic variables are akin to the order parameters used in the traditional theory of phase transitions and can be used in order to detect in a simple way the types of change occurring in multiple systems. Moreover, their introduction allows to define in a simpler way the concept of coherence and other meta-structural properties.

We may now take into consideration a multiple system or a CB constituted, to fix the ideas, by a number \textit{k} of elements $e_j$ (\textit{j}=1,...,\textit{k}). We then denote by $m_r(t_i)$ (\textit{r}=1,...,\textit{s})  one of the eventual \textit{s} mesoscopic variables whose value characterizes the CB at the instant $t_i$. Now the element $e_j$ can or cannot belong to the set of elements contributing to the actual value of $m_r(t_i)$. To take into account such a circumstance we can introduce a further variable $e_{jr}$, assuming the value 1 if at the time $t_i$ the element $e_j$ belongs to the set of elements contributing to the actual value of $m_r(t_i)$, and assuming the value zero in the contrary case. This allows to introduce, for each element, the \textbf{mesoscopic general vector}:

	\[V_j(t_i)\equiv\left\lfloor e_{j1}(t_i),e_{j2}(t_i),...,e_{js}(t_i)\right\rfloor
\]
as well as the \textbf{mesoscopic general matrix} whose elements are $e_{jr}(t_i)$. In order to make shorter the definitions which will appear in the following, we will use the expression ``the element (or agent) $e_{j}$ possesses the mesoscopic property $m_r$, at the time $t_i$'' if and whenever $e_{jr}(t_1)=1$. Though this expression is incorrect, we will use it only as a purely formal statement avoiding the use of too many symbols.\\
The study of meta-structural properties, however, cannot be successfully undergone without the introduction of a further set of variables, each one associated to each mesoscopic variable. We will denote these variables as \textit{parametric variables}. The value of the \textit{r}-th parametric variable associated to the \textit{r}-th mesoscopic variable at the time $t_i$ can be defined as the average value, on the set of all elements such that $e_{jr}(t_i)=1$, of the observed values of the variable characterizing the single elements, from which the mesoscopic variable itself was derived through the coarse graining procedure described above. The value of the parametric variable associated to the \textit{r}-th mesoscopic variable at the time $t_i$   will be denoted by $p_r(t_i)$ .
To make a simple example, let us have a system with only 5 elements defined through a single variable -- for instance their velocity in m/s-- having, at a given time, the observed values defined by the following vector:

	\[\nu\equiv(3,5,1.5,1.25,1).\]
If we introduce a threshold of 0.25 it is easy to see that the 3\textsuperscript{rd}, 4\textsuperscript{th}, and 5\textsuperscript{th} element have the same velocity and the mesoscopic general vector has the form:
\[V=
\left[
\begin{array}{ccc}
0\\
0\\
1\\
1\\
1
\end{array}
\right]
\]
Moreover, the value of this specific mesoscopic variable is 3 (number of elements having the same velocity) while the value of the associated parametric variable is 1.25 (the average of the three velocity values observed for the 3\textsuperscript{rd}, 4\textsuperscript{th}, and 5\textsuperscript{th} element).

Dealing now with the dynamics of a CB, i.e., described by a time sequence of mesoscopic general vectors, it is possible to consider four typical situations.

\begin{enumerate}
	\item \textbf{All the agents simultaneously possess all the same mesoscopic properties, and the associated mesoscopic and parametric variables have values constant through time}. In this case not only 
	$\forall i\forall j\forall r \left\lfloor e_{jr}(t_i)\right\rfloor=1$ but we have also that $\forall r\forall i\forall k\left[m_{r}(t_i)=m_r(t_k)\right]$ and $\forall r\forall i\forall k\left[p_{r}(t_i)=p_r(t_k)\right]$. This case corresponds to a simple collective behavior of all agents rigidly fixed and whence to a trivial case of meta-structures.
	\item \textbf{All the agents simultaneously possess all the same mesoscopic properties, and the associated mesoscopic variables have parametric values changing with time.} The only difference with the previous case is that $\exists r\exists i\exists k \left[p_r(t_i)\neq p_r(t_k)\right]$. In this situation we have only parametric changes of a given meta-structure.
	\item \textbf{The agents possess different mesoscopic properties but the parametric values are \textit{constant} through time.} In this case different elements are associated to different general mesoscopic vectors, even if the parametric values associated to the mesoscopic variables do not change with time. This situation corresponds to a complex pattern of collective behaviors. 
	\item \textbf{The agents possess different mesoscopic properties but the parametric values are changing with time.} This case corresponds to the patterns of collective behaviors characterized by the highest complexity.
	\end{enumerate}

	These four situations are synthetically listed in the following Table \ref{tab:Mesoscopic Dynamics}.
	
	\begin{table}[h]
		\centering
			\begin{tabular}{|p{3.5cm} | p{3.5cm} | p{3.5cm} | p{3.5cm}|}
			\hline
  \multicolumn{4}{|c|}{Mesoscopic Dynamics} \\
				\hline
				\hspace{1.25cm} Case1 & \hspace{1.25cm}  Case2 & \hspace{1.25cm}  Case 3 & \hspace{1.25cm}  Case 4\\
				\hline \textit{All} the agents possess \textit{all} the same mesoscopic properties and parametrical variables \textit{constant} through time (only \textit{insignificant} changes within the threshold assumed) & \textit{All} the agents possess \textit{all} the same mesoscopic properties, but the parametrical variables are changing with time & The agents possess different mesoscopic properties whose parametric values are \textit{constant} through time (only \textit{insignificant} changes within the threshold assumed) & The agents possess different mesoscopic properties whose parametric values are changing with time\\ 
				
\vspace{0.5cm} \textit{Mesoscopic structure fixed} & \vspace{0.5cm} \textit{Changes of the same mesoscopic structure} & \vspace{0.5cm} \textit{Multiple and superimposed implementations of the same mesoscopic structures} & \vspace{0.5cm} \textit{Multiple and superimposed variations of the mesoscopic structures}\\
\hline
 Trivial meta-structural properties & Trivial meta-structural properties & Non-trivial meta-structural properties & Non-trivial meta-structural properties 
\\
\vspace{0.5cm} \textit{Collective behaviours structurally `fixed'} &  \vspace{0.5cm} \textit{Collective behaviours structurally at low variability }  & \vspace{0.5cm} \textit{Collective behaviours structurally variable} & \vspace{0.5cm} \textit{ Collective behaviours structurally at high variability }\\ \hline
\multicolumn{4}{c}{}\\
\multicolumn{4}{c}{----------------------------------------------------------------------------------------------------------------$>$}\\
\multicolumn{4}{c}{Direction representing increasing of complexity due to increasing of structural change}\\
				
			\end{tabular}
		\caption{Mesoscopic Dynamics}
		\label{tab:Mesoscopic Dynamics}
	\end{table}
	
	\section{Meta-structural properties}
	The meta-structural properties, i.e., the mathematical properties of the sets of values assumed by mesoscopic as well as parametrical variables over time, can be used to characterize the \textit{coherenc}e of the mesoscopic dynamics. The latter can be detected through the coherence of sequences of multiple and superimposed structures, which in turn produces microscopic coherence at the level of the single elements.
	
Examples of meta-structural properties are:

\begin{itemize}
	\item Eventual kinds of \textit{regularity} of the values acquired by mesoscopic and parametric variables, like periodicity, quasi-periodicity and possible chaotic behaviour establishing attractors characterising specific CBs;
	\item Eventual \textit{cross-correlations} between variables;
	\item Occurrence of invariant or of transitions detected through the methods of computational statistical mechanics;
	\item Eventual statistical properties detected through techniques like:
	\begin{itemize}
		\item Multivariate Data Analysis (MDA) and Cluster Analysis;
		\item Principal Component Analysis (PCA);
\item Recurrence Quantification Analysis (RQA);
\item Time-Series Analysis.
	\end{itemize}
	\end{itemize}
	
	\section{The Meta-structures project}
	We now introduce some considerations on the implementation of the project devoted to investigate Meta-Structural properties in CBs (see, in this regard, the project web site \cite{web1}). 
	
The first stage of research concerns \textit{simulated} CBs, at a suitable level of complexity, in that they make all necessary microscopic values available and it is possible to act on the threshold and parametric values. This simplified context of research has been introduced to test initial approaches to be subsequently reconsidered since meta-structural properties deal with phenomena having properties \textit{irreducible} to rules, while computational emergence is given by rules. 
A first version of the software for the simulation of CBs with significant complexity, which can be described at a microscopic level suitable to set up mesoscopic variables as introduced above, is available at the project web site \cite{web2}. The simulation software is able to give all the microscopic information necessary to establish mesoscopic variables. Any output is MATLAB compatible.

The first phase of the research will take place within a \textit{simplified} context given by computational emergence. This context constitutes the \textit{computational laboratory} to search for meta-structural properties.

The eventual success in finding meta-structural properties within this simplified context is considered methodologically useful before applying the approach to cases of non-simulated emergence such as industrial districts, markets, traffic, urban development (morphology; energy behaviour; induction of behaviour to agents inhabiting structures) and ecosystems, for which all microscopic data are available. We mention an experimental computational approach based on varying a simulated CB by inserting within it another suitable eventually adaptive CB as \textit{order parameter} \cite{web3}. Different approaches are possible such as when some agents assume at a defined time a new collective interaction by a) setting the eventual varying distribution within the CB; b) varying the number of agents assuming a new CB; c) using a learning mechanism.

\section{Some research issues}
An interesting research issue is the eventual relationship between the four classes of \textit{mesoscopic dynamics} and the four classes of cellular automata introduced by Wolfram (see, for instance, Wolfram, 1994 \cite{Wolfram1994}):

\begin{itemize}
	\item Class 1: evolve into stable, homogeneous structures. 
	\item Class 2: evolve into stable or oscillating structures. Local randomness.
	\item Class 3: chaotic evolution. Spread randomness.
	\item Class 4: emergence of local and surviving dynamic structures.
	\end{itemize}
	
Another important line of research concerns the relationships between meta-structures and complex networks (see, for instance, Lewis, 2009 \cite{Lewis2009}; and also Motter and R\'eka, 2012 \cite{Motter2012}; Valente, T. W. 2012 \cite{Valente2012}). Meta-structural properties of complex networks represented by properties of suitable mesoscopic variables such as, for instance, \textit{topological distances} between nodes, families of links, of nodes, and of \textit{fitness} could be taken into account to detect, induce, keep or restore \textit{multiple}, \textit{even} superimposed, \textit{topological properties and topological behaviours of complex networks}. 

\section{Conclusions}

In this paper we presented, having modellers and designers in mind, some theoretical notes and approaches based on considering CBs as ruled by a coherent mesoscopic dynamics. The paper contains specifications useful for researchers interested in considering meta-structures as a conceptual framework to model phenomena related to CBs. The same specifications can eventually be useful to design CBs by both a) directly and explicitly \textit{prescribing} mesoscopic coherence and b) \textit{inducing} and \textit{varying} emergence of CBs when prescribing meta-structural properties.

\nocite{*}
\bibliographystyle{eptcs}
\bibliography{generic}

\textbf{Acknowledgments}-Development of the simulation software at \cite{web1} has been supported by CSE (Corporate Strategy Expertise)-CRESCENDO, Milan, Italy http://www.cse-crescendo.it/
\end{document}